\documentclass[preprint]{vgtc}               

\ifpdf
  \pdfoutput=1\relax                   
  \usepackage{graphicx}                
  \DeclareGraphicsExtensions{.pdf,.png,.jpg,.jpeg} 
\else
  \usepackage{graphicx}                
  \DeclareGraphicsExtensions{.eps}     
\fi%

\graphicspath{{figures/}{pictures/}{images/}{./}} 

\usepackage{microtype}                 
\PassOptionsToPackage{warn}{textcomp}  
\usepackage{textcomp}                  
\usepackage{mathptmx}                  
\usepackage{times}                     
\usepackage{cite}                      
\usepackage{tabu}                      
\usepackage{booktabs}                  

\usepackage[table,xcdraw,x11names,dvipsnames]{xcolor}
\usepackage[hypcap=false]{caption}
\usepackage{subcaption}
\usepackage{enumitem}

\usepackage{multirow}
\usepackage{graphicx}

\newcommand{\whiteboard}{\emph{Whiteboard} }
\newcommand{\clustervis}{\emph{ClusterVis} }

\onlineid{0}

\vgtccategory{Research}

\usepackage{tikz}
\usepackage{textcomp}
\usepackage{hyperref}
\usepackage{lipsum}

\newcommand\copyrighttextmy{%
  \footnotesize \textcopyright 2022 IEEE. Personal use of this material is permitted. Permission from IEEE must be obtained for all other uses, in any current or future media, including reprinting/republishing this material for advertising or promotional purposes, creating new collective works, for resale or redistribution to servers or lists, or reuse of any copyrighted component of this work in other works.
  }
\newcommand\copyrightnotice{%
\begin{tikzpicture}[remember picture,overlay]
\node[anchor=south,yshift=10pt] at (current page.south) {\fbox{\parbox{\dimexpr\textwidth-\fboxsep-\fboxrule\relax}{\copyrighttextmy}}};
\end{tikzpicture}%
}

\title{Visual Firewall Log Analysis - At the Border Between Analytical and Appealing}

\author{Marija Schufrin \thanks{The first two authors contributed equally to this work.} \thanks{e-mail: marija.schufrin@igd.fraunhofer.de}\\ %
        \parbox{1.4in}{\scriptsize \centering Fraunhofer IGD \\ TU Darmstadt} %
\and Hendrik Lücke-Tieke \footnotemark[1] \thanks{e-mail: hendrik.luecke-tieke@igd.fraunhofer.de} \\ %
     \scriptsize Fraunhofer IGD %
\and  Jörn Kohlhammer \thanks{e-mail: joern.kohlhammer@igd.fraunhofer.de}\\ %
     \parbox{1.4in}{\scriptsize \centering Fraunhofer IGD \\ TU Darmstadt}
     }

\teaser{
    \includegraphics[width=\linewidth]{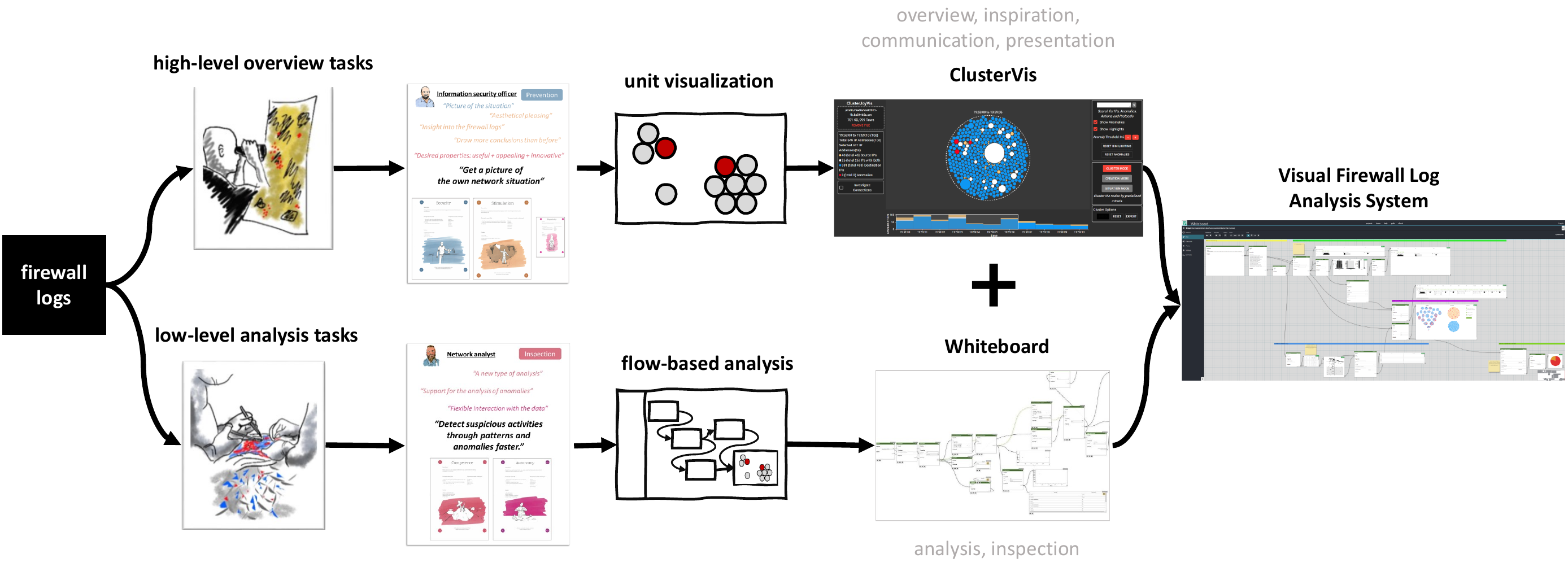}
  \caption{ 
    With the goal to provide insights into the firewall logs of an organization, we identified two types of interests: 
  high-level overview and low-level analysis. A persona with the main targeted psychological needs was defined for each role (information security officer and network analyst).  
  We developed two concepts to fit to the requirements for each usage context and two interfaces 
   that can be used together as a combined visual firewall log analysis system. 
  (Need cards \copyright Hassenzahl et al. \cite{hassenzahlExperienceDesignTools})
  }
}

\abstract{In this paper, we present our design study on developing an interactive visual firewall log analysis system in collaboration with an IT service provider. 
We describe the human-centered design process, in which we additionally considered hedonic qualities by including the usage of personas, psychological need cards and interaction vocabulary.
For the problem characterization we especially focus on the demands of the two main clusters of requirements: high-level overview and low-level analysis, represented by the two defined personas, namely information security officer and network analyst. 
This resulted in the prototype of a visual analysis system consisting of two interlinked parts. One part addresses the needs for rather strategical tasks while also fulfilling the need for an appealing appearance and interaction. The other part rather addresses the requirements for operational tasks and aims to provide a high level of flexibility.
We describe our design journey, the derived domain tasks and task abstractions as well as our visual design decisions, and present our final prototypes based on a usage scenario. 
We also report on our capstone event, where we conducted an observed experiment and collected feedback from the information security officer.
Finally, as a reflection, we propose the extension of a widely used design study process with a track for an additional focus on hedonic qualities.%
} 

\CCScatlist{
  \CCScatTwelve{Human-centered computing}{Visu\-al\-iza\-tion}{Information visualization}{}; \CCScatTwelve{Security and privacy}{Human and societal aspects of security and privacy}{}{};
}

\begin{document}


\firstsection{Introduction}

\maketitle
\copyrightnotice

A strong network security is an indispensable requirement for all organizations with an IT infrastructure. 
This is especially true for IT service providers. 
In this context, effective methods to support the routine tasks of the responsible persons can have a valuable impact on the organizations network security \cite{bottaUnderstandingITSecurity2007}. 
Applying visualization methods to increase the visibility of different logs gathered from the IT network can increase the level of insight and lead to better decisions \cite{shiraviSurveyVisualizationSystems2012}. 
A firewall provides an important perimeter for the network that also allows the observation of incoming and outgoing traffic. This traffic (accepted and denied connections) is recorded in firewall logs that may contain valuable information concerning the activities on and around the network \cite{martyAppliedSecurityVisualization2009,bondVisualizingFirewallLog2009}. 
The visual analysis of firewall logs was the main goal of a joint project with an IT service provider.
During our multi-year design study \cite{sedlmair2012design}, we have encountered a number of challenges, which we present in this paper together with our proposed solutions.

The first challenge was the often-encountered problem to balance conflicting requirements of several stakeholders.
We show how we have identified two clusters of users, represented each of them by a persona \cite{miaskiewicz2011personas} and designed the joint solution with respect to these requirements.
Our proposed system approach interlinks two interfaces in one solution, while addressing the individual needs of the user groups with each interface.
Our second challenge was to find appropriate visual solutions to enable flexible and interactive insights from firewall log analysis. 
Our solution combines an analytical interface with interactive overview visualizations.
The third challenge was to adequately integrate non-pragmatic (or hedonic) aspects into our design process.
The need for this emerged as our collaborators, beside the functional requirements, expressed the desire for a notably positive and aesthetic appeal of the final solution. 
Researchers in HCI emphasize the role of considering subjectively perceived qualities for the design of products and software since many years \cite{hassenzahl2000hedonic}.
Hassenzahl et al.\cite{hassenzahl2000hedonic} coined the term \textit{hedonic} as opposed to pragmatic qualities. 
While pragmatic qualities comprise quality dimensions that
are related to traditional usability and focus on task-related functions or
design issues (targeting the so called \textit{do-goals}), "hedonic qualities comprise quality dimensions with no obvious relation to the task the user wants to accomplish with the system, such as originality, innovativeness, beauty" \cite{hassenzahl2000hedonic}, addressing the so-called \textit{be-goals}. 
Hassenzahl et al. propose to consider basic psychological needs to achieve hedonic qualities \cite{hassenzahlNeedsAffectInteractive2010}.
In this paper we describe how we used this approach in our design study and finally propose a process model to include this approach into the infovis design study pipeline.
In particular, we considered psychological needs \cite{hassenzahlNeedsAffectInteractive2010} and interaction vocabulary \cite{lenz2013exploring}. %
Along the lines of Sedlmair et al. \cite{sedlmair2012design}, the three contributions of this paper are:
\begin{enumerate}[noitemsep]
    \item \textbf{Problem characterization and abstraction:} Domain characterization for visual firewall log analysis in IT organizations for multiple decision makers with strongly different requirements
    \item \textbf{Validated visualization solution:} Proposed concept of a visualization system consisting of two interlinked parts, considering pragmatic as well as hedonic aspects, which is implemented and validated as a web-based prototype
    \item \textbf{Reflection:} Proposition of an extended design process model taking into account hedonic qualities by including personas and the psychological needs theory
\end{enumerate}

\section{Related Work}
Three topics are relevant for the related work: existing visualization approaches for IT network logs, visual environments based on data flows, and work addressing pragmatic and hedonic design choices. 

\textbf{Visualizations for network log analysis}
There are already many visual analytics approaches for the analysis of network logs (see surveys from Shiravi et al. \cite{shiraviSurveyVisualizationSystems2012} and Zhang et al. \cite{zhang2012survey, zhangSurveyNetworkAnomaly2017}). 
There are approaches that only focus on the interactive visualization of these logs \cite{ma2004visualization, leeVisualFirewallRealtime2005, abdullah2005ids, fischerClockMapEnhancingCircular2012}, 
but also approaches that incorporate automatic data processing and detection in some way \cite{matsumoto2010method, gove2021automatic, guerra2019astudy}.
Several publications focus on one visualization technique \cite{ballHomecentricVisualizationNetwork2004, choiFastDetectionVisualization2009, yinVisFlowConnectNetflowVisualizations2004}, while others combine various visualizations and different views
\cite{cappers2015snaps, goodallPreservingBigPicture2005a, ghoniemVAFLEVisualAnalytics2014,arendtCyberPetriCDX20162016}.
Further, some approaches focus only on one log type (e.g. firewall log data) \cite{ghoniemVAFLEVisualAnalytics2014, matsumoto2010method}, while others aim to combine logs from different data sources
\cite{freiHistogramMatrixLog2008, gove2021automatic, fischerLargeScaleNetworkMonitoring2008} or existing information about known malware or attacks.
The most frequently used visualization techniques we observed were \textit{node-link diagrams} \cite{arendtOcelotUsercenteredDesign2015,bond2009visualizing}, \textit{matrices} \cite{freiHistogramMatrixLog2008, ghoniemVAFLEVisualAnalytics2014}, and \textit{parallel coordinates} \cite{yinVisFlowConnectNetflowVisualizations2004, choiFastDetectionVisualization2009}. Also \textit{symbol or glyph based visualizations} \cite{komlodi2005auser, fischerClockMapEnhancingCircular2012} are present. 
There are also many approaches with hierarchical layouts.
These range from binary arrangement (internal vs external) \cite{ballHomecentricVisualizationNetwork2004, fischerLargeScaleNetworkMonitoring2008} to rectangular \cite{fischerLargeScaleNetworkMonitoring2008} and circular \cite{fischerClockMapEnhancingCircular2012, arendtOcelotUsercenteredDesign2015,arendtCyberPetriCDX20162016} tree maps.
Apart from hierarchical layouts, overlays are frequently applied to visualize the actual communication behaviour. For example, edges are drawn between sources and targets to show the detailed connections on top of the hierarchical context
\cite{ballHomecentricVisualizationNetwork2004, fischerLargeScaleNetworkMonitoring2008, kintzelMonitoringLargeIP2011, arendtCyberPetriCDX20162016}, as we do in \textit{ClusterVis}.
Several cyber-security visualizations use static or animated bubbles to convey information ranging from global scale down to the individual packet \cite{mckennaBubbleNetCyberSecurity2016, arendtOcelotUsercenteredDesign2015, arendtCyberPetriCDX20162016,komlodi2005auser}.
There are also several approaches that we perceived as appealing or pleasing to the eye, such as 
\cite{fischerClockMapEnhancingCircular2012},
\cite{mckennaBubbleNetCyberSecurity2016} or \cite{chen2012semanticprism}.
In summary, while there are approaches for network log analysis that are well suited for their goals of which some provide an appealing appearance, most of them have limited flexibility. Therefore, we added the concept of visual flow-based data analysis to our design solution. 
We have not found an approach for network log analysis that uses a visual flow-based data analysis in combination with an interactive visualization for the analysis of network log data.

\textbf{Flow-based data processing and analysis}
Visual environments based on data flows are well suited to allow users to flexibly and interactively choose analytical methods to process the firewall log data.
Several approaches for general data analysis exist, such as KNIME \cite{bertholdKNIMEKonstanzInformation2008} or YALE \cite{mierswaYALERapidPrototyping2006}. Lately, more interactive elements are embedded directly into the workflow \cite{yuVisFlowWebbasedVisualization2017, demsarOrangeDataMining2013}.
Although visual programming environments have strongly advanced in recent years and enable users to solve problems computationally without the need of learning programming languages, they are mostly suited for data analysts. Thus, one valuable extension is to enrich such environments with attractive and interactive visualizations.
For example, VisFlow \cite{yuVisFlowWebbasedVisualization2017} integrates interactive visualizations directly on the canvas. We have used this approach in the analytical \textit{Whiteboard}, which especially allows to integrate the more appealing \clustervis and other visualizations.
Additionally, we feature a full data-flow processing model instead of data subset flows. 

\begin{figure*}
        \includegraphics[width=\linewidth]{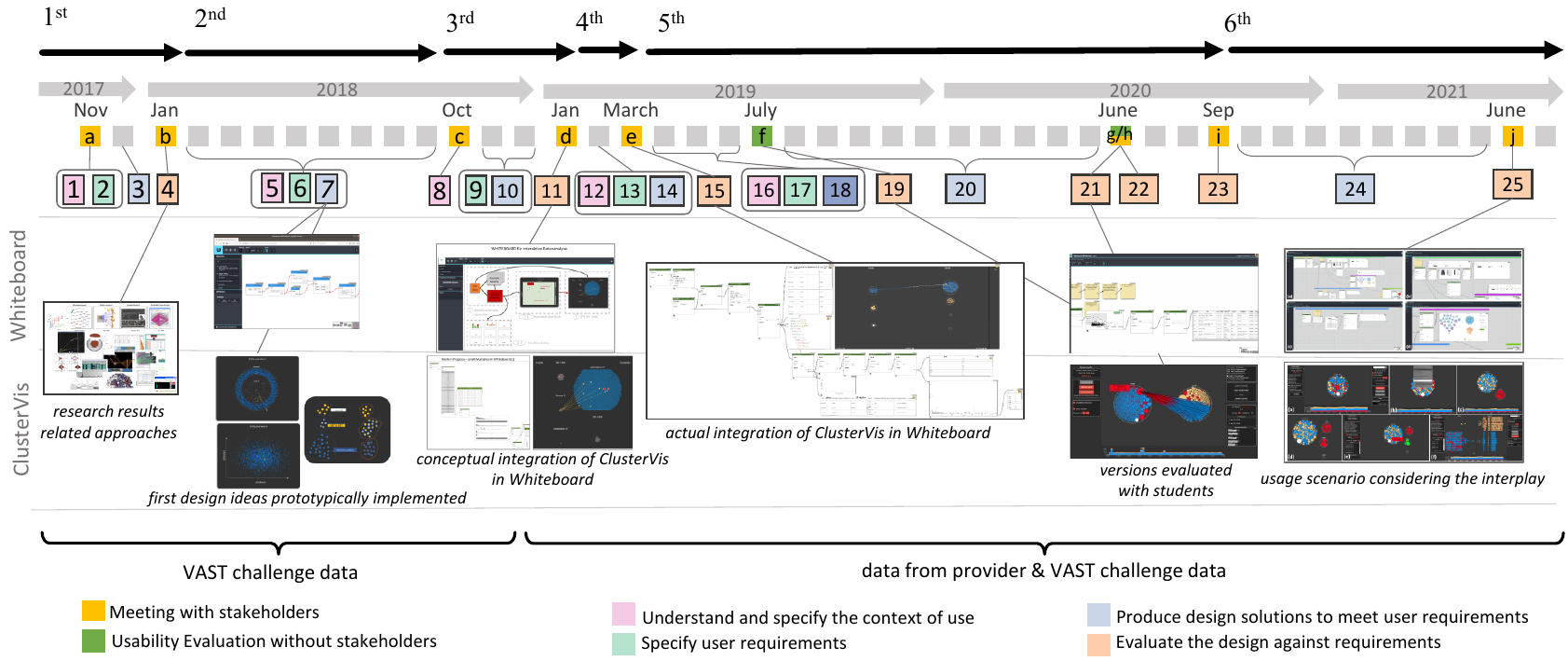}
       \caption{
       The journal of events of our design study with six iterations. The timeline shows the events with other participants (yellow and green) and milestones according to the four steps of the HCD cycle \cite{ISO9241-UCD}. Selected screenshots at particular events show the parallel development of the two prototypes \whiteboard and \clustervis.
       }
    \label{fig:dev_journey}
\end{figure*}
\textbf{Pragmatic and hedonic visualization design}
Along the lines of the human-centered design process (HCD) \cite{ISO9241-UCD} and the research of Hassenzahl et al. \cite{hassenzahl2010experience}, in our design process we tried to consider both pragmatic and hedonic qualities.
Looking at established process models for infovis and visual analytics design, we observed a lack of guidance with regard to hedonic qualities.
For example, while in the visualization pipeline of Card and Mackinlay \cite{card1999readings} the user is an integral part of the process of the interactive transformation of data into visual forms, the focus is on the pragmatic, visualization-related aspects.
The same is true for the design triangle of Miksch and Aigner \cite{mikschMatterTimeApplying2014}: while they characterize user needs along various axes, the main focus is on pragmatic qualities.
Munzner's Nested Model \cite{munznerNestedModelVisualization2009} as well as the methodology proposed by Sedlmair et al. \cite{sedlmair2012design} provide frameworks for design studies in information visualization. While they leave room to integrate hedonic qualities into each step, these potential choices are not explicitly mentioned.
Several authors promote a stronger interplay of art and information visualization \cite{lau2007towards,moere2011role,thudt2012bohemian,samsel2018art}, while others stress the importance of engagement \cite{hung2017assessing, mahyar2015towards} and user experience \cite{schufrintowards2018}.
However, we have not found related work that actually reports on such methods in their visualization design process.
Thus, to the best of our knowledge, there is currently no model that gives guidance on integrating both pragmatic and hedonic design choices in visualization design study.

\section{Our Development Journey} 
This section gives an overview of the development process, highlights the key events, and outlines the challenges we had to overcome (see \autoref{fig:dev_journey}). 
Our journey took around three and a half years and involved several iterations. 
The \textbf{first HCD iteration} started with the goal to interactively investigate data from the cyber-security domain. A first meeting (a) with an IT service provider revealed the demand for a visual security solution (1). More precisely there was a requirement for analytics of available internal logs (2). Based on that we foraged for related visualization approaches for network security (3) and presented them in a meeting (b) to our partners at the IT provider (4). 
Summarizing the findings within the \textbf{second HCD iteration}, we observed that there are various stakeholders with different interests in the context of network data analysis (5). Further, the main interest was to \textit{provide visual insights into internal network logs} of the organization with a focus on perimeter firewall logs (6).  
We started with baseline work on a research prototype (7) based on our initial understanding with several updates during our journey.
Thereby, we worked both on the analytical part - \textit {which analysis could be applied to firewall logs} (\textit{Whiteboard}) and on the visual part, \textit{how the content of firewall logs can be visualized} (which later resulted in \textit{ClusterVis}).
Initially, we used the VAST challenge 2012 dataset \cite{vast2012}.
The \textbf{third HCD iteration} started with a meeting with the information security officer (c), where we specified the context of current technical possibilities, infrastructure, data and possible targeted users and goals (8). Based on that we derived a first description of the requirements and the relevant data, users and tasks (9).  
Our collaborators also emphasized non-functional requirements in the direction of "aesthetically pleasing" elements, literally asking for "eye candy" to complement the more analytics-oriented approaches.
Therefore, we integrated user experience (UX) techniques into our design process. 
At this stage, we employed personas \cite{changPersonasTheoryPractices2008} and considered the psychological need cards \cite{hassenzahlExperienceDesignTools}. 
We identified two main groups of stakeholders: technical experts and non-technical staff.
Thus, we decided to proceed with the parallel development of two prototypes - one with a stronger analytical and one with a stronger visual focus. 
These two prototypes were planned to be interlinked with each other by integrating the \textit{ClusterVis} visualization into the analytical \textit{Whiteboard}.
We concluded the third iteration by presenting the prototypes to the information security officer in a follow-up meeting (d) and received an anonymized sample of the real record of the firewall logs of this organization. 
We used the findings for the \textbf{fourth HCD iteration} to fine-tune our understanding of the targeted users, involved data and tasks (12) as well as the requirements (13). We also improved the prototypes on this basis and adapted them to the new data (14).
The highlight of this iteration was a workshop (e) with the information security officer, a network analyst and a firewall expert. Focusing on the interests and feedback of the network analyst, we could collect more detailed information about actual daily tasks, fine-tune our requirements, and receive feedback on the current version (15).
This feedback and an improved understanding of the needs and use cases of the potential stakeholders directly influenced the \textbf{fifth HCD iteration}, 
which consisted of improvements of our descriptions of the context of use (16), the requirements (17) and the prototypes (18). 
At this stage we also applied the interaction vocabulary \cite{lenz2013exploring, hassenzahlExperienceDesignTools} to sharpen the interaction design of the prototypes.
Further we conducted a usability evaluation of \whiteboard (f) (19) with n=13 students and of \clustervis (g)(21) with n=31 students, and presented the results to related stakeholders (h)(22). We concluded with a feedback meeting with the information security officer (i)(23). Meanwhile, we continuously worked on the prototypes (20). 
The results of the evaluation helped fix many usability issues of both prototypes.
At the end of the \textbf{sixth HCD iteration}, after further prototype improvements (24), we conducted a capstone event with feedback session and final evaluation (25) with the information security officer.

\section{Problem Characterization and Abstraction}\label{sec:problem_characterization}
\begin{figure} [h!]
    \includegraphics[width=\linewidth]{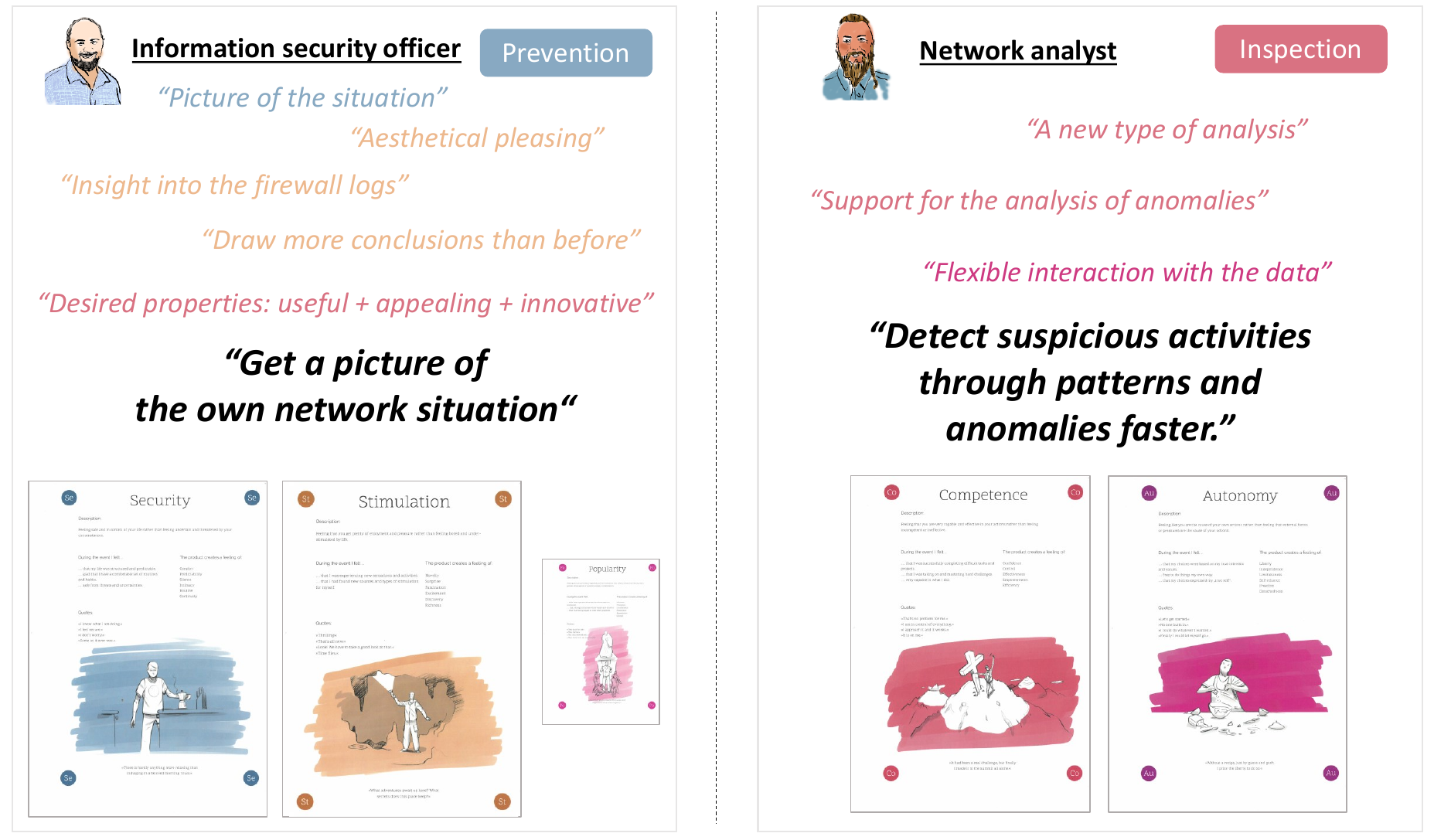}
    \caption{Two personas: the information security officer requires an appealing and fast high-level overview of the network situation. The network analyst is interested in a faster detection of suspicious activities. To support the design process we have assigned psychological needs to each persona. (Need cards \copyright  Hassenzahl et al. \cite{hassenzahlExperienceDesignTools})}  
    \label{fig:personas}
\end{figure}

We identified the demand to \textit{get visual insights into internal network logs of the organization} as the main point of interest for our collaboration. However, during the first interviews we observed that different potential stakeholders and interests are involved.
After multiple interviews, we identified two prominent clusters of requirements from different stakeholders with intersecting interests (see \autoref{fig:table_problem_cluster}).
Two main global usage scenarios emerged: On the one hand, there was the need for \textit{high-level overview} tasks and a focus on \textit{prevention} and \textit{reporting}, including an appealing appearance. On the other hand, there was a demand for \textit{low-level analysis} during daily routine with a focus on \textit{inspection} and \textit{detection}. 
In this paper, Cluster A is represented by the persona \textit{information security officer} and Cluster B by the persona \textit{network analyst} (see \autoref{fig:personas}).

\begin{table}[h!]
 \caption{Selected characteristics of the two identified clusters. 
 }  
\includegraphics[width=\linewidth]{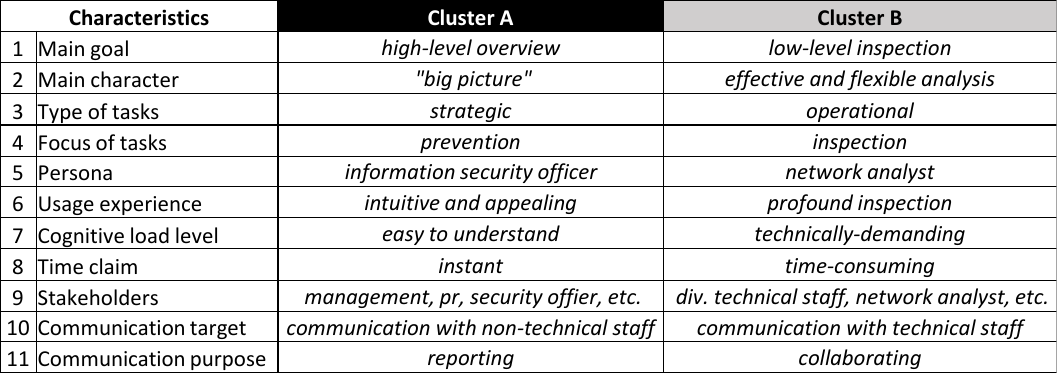}
 \label{fig:table_problem_cluster}
\end{table}

\begin{figure} [t!]
     \includegraphics[width=\linewidth]{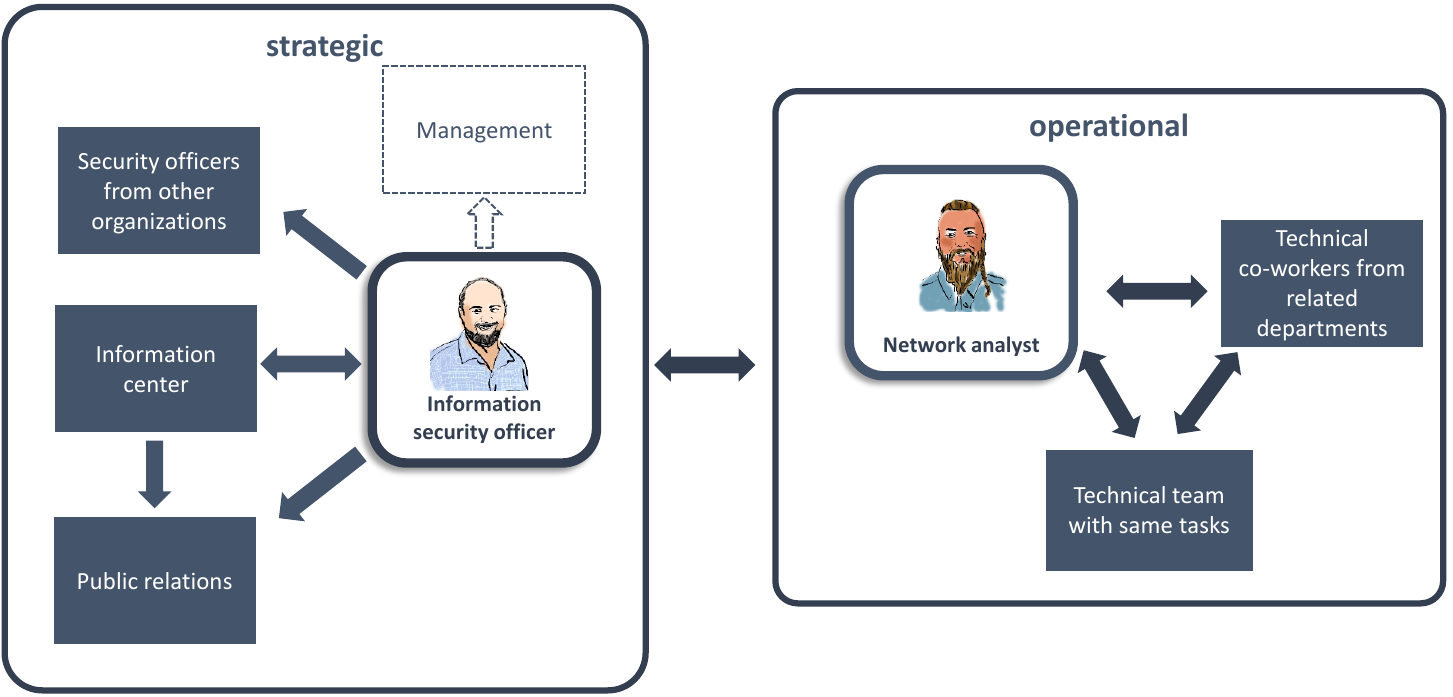}
       \caption{Communication relationships: Information security officer - many communications to persons with a low technical level. Network analyst - mostly regards technical-savvy communication partners. The communication between information security officer and network analysis expert is of particular relevance.}
      \label{fig:communication}
  \end{figure}

\textbf{Users} \label{sec:users}
The research process revealed a wide range of potential stakeholders for the visualization solution we are targeting. With regard to the two identified usage scenarios we decided to represent each cluster by one persona \cite{changPersonasTheoryPractices2008}. 
In \autoref{fig:personas} we abstractly present the two personas \textit{information security officer} as a representative for the staff with a potentially more strategic point of view, and the \textit{network analyst} as a representative for the operational staff. The selection of these personas was also nicely related to our main contacts at our collaboration partner, but is also representative for the broader group of stakeholders.
It is clear, that the interests of the two personas partly overlap. 
As an additional method, we decided to consider the dimension of psychological needs to support our design decisions during our design process.
In the course of this, we used the design cards of Hassenzahl et al. \cite{hassenzahlExperienceDesignTools}.
Based on the interviews, we selected the most appropriate psychological needs and assigned them to the personas.
For the information security officer we chose \textbf{security} and \textbf{stimulation} as the primary needs, and additionally \textbf{popularity}. Security is described as \textit{"feeling safe and in control of your life"} and stimulation as \textit{"feeling that you get plenty of enjoyment and pleasure"} \cite{hassenzahlExperienceDesignTools}. 
For the network analyst, we focused on the need for \textbf{competence} and \textbf{autonomy}. Competence supports a \textit{"feeling that you are very capable and effective in your actions"} and autonomy a \textit{"feeling that you are the cause of your own actions"} \cite{hassenzahlExperienceDesignTools}.
\autoref{fig:personas} shows the assigned cards.
We provide more details on how this inspired our approach and gave us orientation for our design decisions in \autoref{sec:vis_design}. 
Another aspect to highlight is the requirement to communicate findings in the network log between different stakeholders. In \autoref{fig:communication}, we have summarized some of the identified communication needs of both information security officer and network analyst.
Note that there are communication relationships both within and between the two clusters. 

\textbf{Requirements and Tasks} \label{sec:requirements_tasks}
\begin{figure*}[t]
  \centering
     \includegraphics[width=\linewidth]{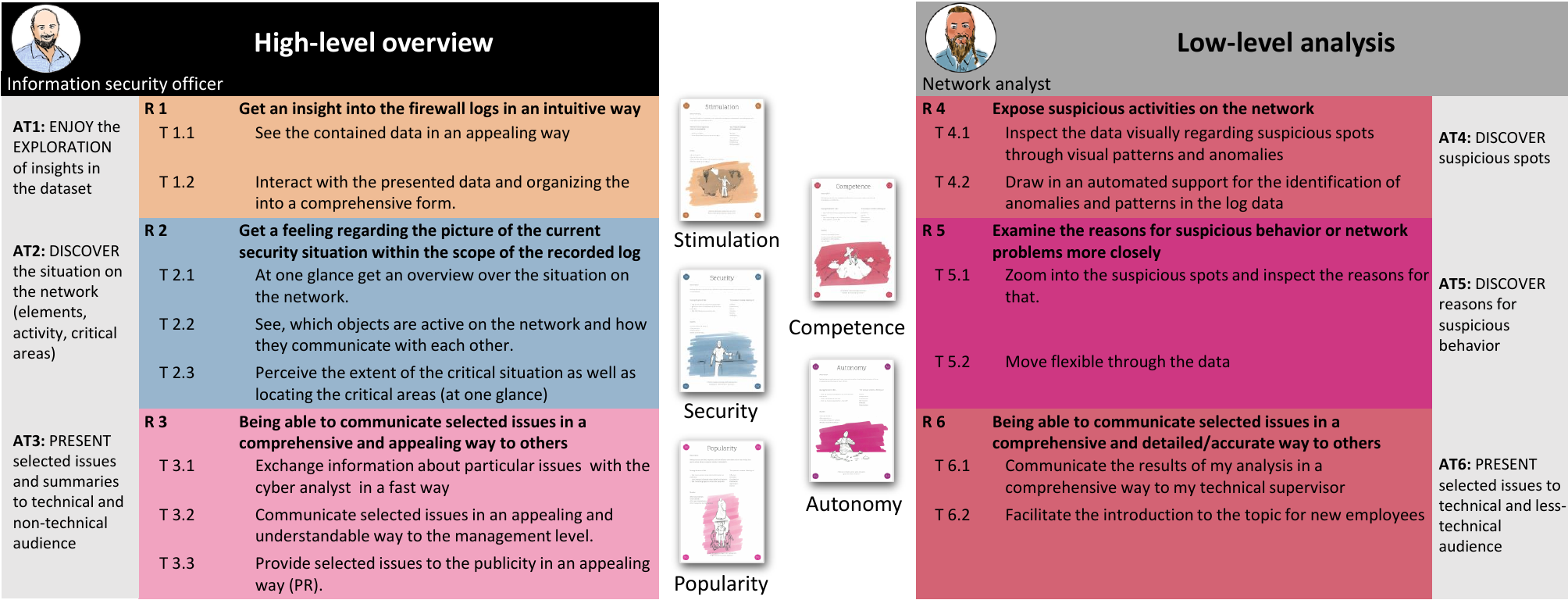}
        \caption{Requirements (R), domain tasks (T) and abstract tasks (AT) resulting from the two main usage areas: high-level overview and low-level analysis representend by the  personas information security officer and network analyst. Requirements and tasks are linked to the psychological needs
      and are translated into abstract tasks according to the taxonomy of Munzner \cite{munzner2014visualization}. (Need cards \copyright   Hassenzahl et al. \cite{hassenzahlExperienceDesignTools})
    } 
    \label{fig:requirements_tasks}
\end{figure*}
In \autoref{fig:requirements_tasks} we summarize the requirements \textbf{R1-R6} and tasks \textbf{T1.1-T6.2} that we identified based on the exchange with our stakeholders. We have defined the six requirements from the perspective of our two personas and assigned each requirement to exactly one persona. While the information security officer (high-level activities) focuses on getting an intuitive overall impression of the current situation in the presented log, the tasks of the network analysis expert (low-level activities) focus on a deep and efficient inspection of the log data. 
However in reality, representatives of both user groups can be interested in each of the task and can benefit from both interfaces. This is also what our final evaluation reveals (see \autoref{sec:evaluation}). 
For each requirement, the list contains a selection of domain tasks, which are defined based on the identified requirements. 
From an infovis perspective, we have translated the six domain requirements to six abstract tasks (\textbf{AT1-AT6}) according to the taxonomy of Munzner et al. \cite{munzner2014visualization,brehmer2013multi} (see \autoref{fig:requirements_tasks} and the supplemental materials for further details). 
Additionally, we assigned the requirements to the psychological needs which is represented by color in \autoref{fig:requirements_tasks}. While there is more than one affiliation (e.g., the communication aspect \textbf{R3} also addresses the need of \textit{competence}), we have focused on the most prominent property for more clarity. 

\textbf{Data} \label{sec:data} 
The data of interest are firewall logs that our project partner stores in large amounts and wants to analyze in a faster and better way.
We received an anonymized example record by the organization, which was recorded at the perimeter firewall and converted into csv format. The log contained more than 100 attributes, of which many were only present under certain conditions.
Our solution is generally applicable to firewall logs, which mostly have a similar basic structure as described in \textit{data abstraction}.
In this collaboration, csv was used, but pcap is supported, too. Also other parsers can be added with low effort.
For initial experiments, e.g., we also used the data set of the VAST challenge 2012 \cite{vast2012}.

\textit{Data Abstraction:} Firewall logs usually contains a lot of relevant information, because all incoming and outgoing traffic has to pass the firewall. Recording both accepted and rejected connections thus leads to a valuable data set. Each log entry contains information about one connection between two IP addresses (inside and outside).
\textbf{Dataset type:} The data set type is primarily \textit{tabular} \cite{munzner2014visualization}. However, as each activity describes the communication of two IP addresses, it can be also viewed as a \textit{network}. 
\textbf{Attribute types:} 
Each log entry includes at least the timestamp (\textit{ordinal}), the source and destination addresses (\textit{categorical}), and the performed actions of the captured activity (\textit{categorical}). 
Depending on the firewall, the log can contain further attributes with different data types, for example, port (\textit{categorical/ordinal}) or protocol (\textit{categorical}).
\textbf{Amount of items:} The amount of items in the data set strongly depends on the activity on the network and on the amount of devices connected to the firewall. The exemplary data record of 10 minutes included around 1.5 million lines. However, in our solution we primarily focused on subsets of around \textit{1.000 to 50.000 entries}. 

\section{Visualization Design} \label{sec:vis_design}
In this section, we present our two resulting inter-playing interfaces and explain our design decisions based on the findings presented in  \autoref{sec:problem_characterization}.
To support our two personas, we designed two interlinked prototypes: A flexible analytical tool (\textit{Whiteboard}) and an interactive visualization with a particular focus on visual appeal (\textit{ClusterVis}). 
The two prototypes can be used stand-alone, but can also be combined by either embedding \clustervis in \whiteboard or by exchanging data-exports \textbf{T3.1}. This approach balances the need for independence and collaboration. 
\subsection{Decisions for ClusterVis (High Level)}
\begin{figure*}[t]
  \centering
     \includegraphics[width=0.95\linewidth]{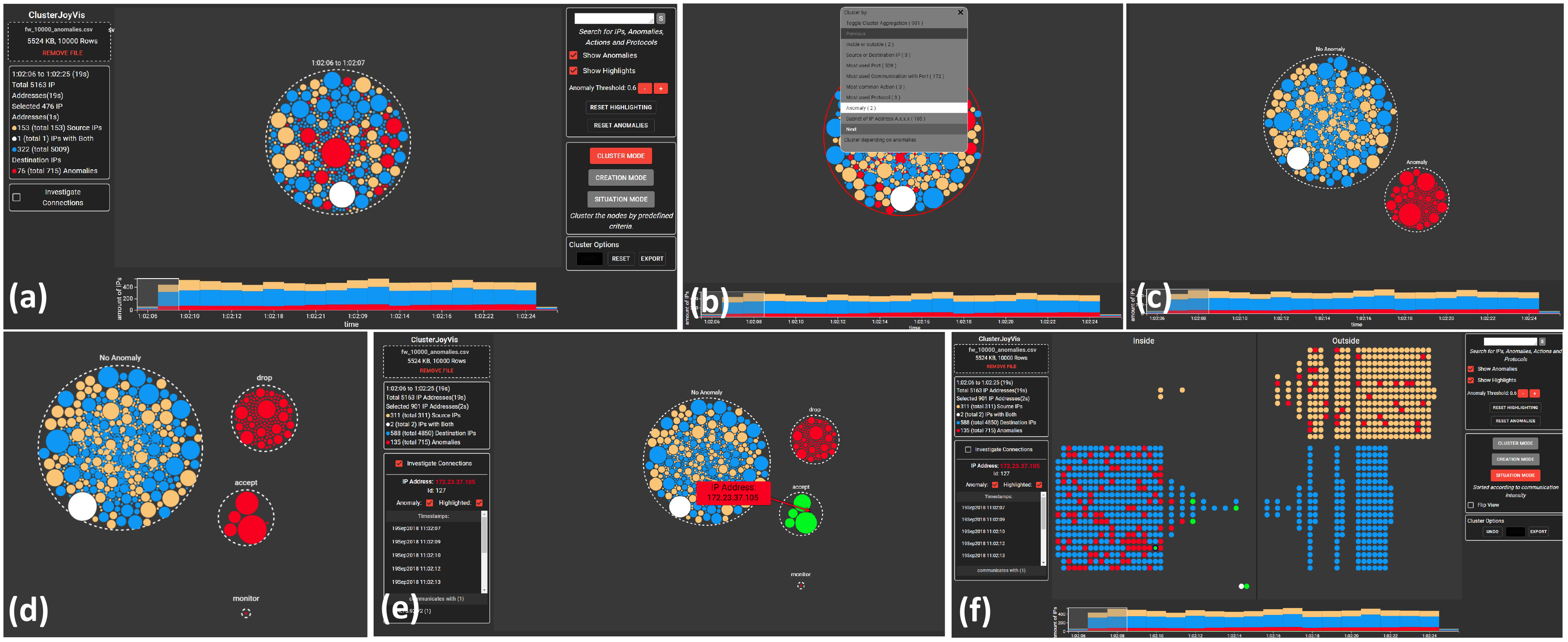}
      \caption{Usage scenario - \textit{ClusterVis}: (a) The data export from \whiteboard is displayed; contained IP addresses are shown as bubbles; anomalous addresses are marked in red. (b) Clicking on a cluster reveals a menu, where the attribute "anomalous" is chosen for a further split. (c) The anomalous IP addresses are now in a separate cluster. (d) The anomalous cluster is further split based on the attribute "most common action", revealing three clusters.
      (e) The accepted anomalies are of particular interest and were brushed green by the user. (f) In situation mode, all IP addresses are arranged near the perimeter. IP addresses with more connections to the other side are closer to the perimeter.
     }    
      \label{fig:use_case_1_clustervis}
  \end{figure*}

\clustervis (see \autoref{fig:use_case_1_clustervis}) is designed for the requirements of the \textit{information security officer} (see \autoref{fig:requirements_tasks}).
The recorded firewall log, in csv format, can be dragged and dropped directly into the interface. The visualization shows each IP address from
 the log as a separate filled circle collected in a cluster, represented by the dotted lines. The main functionality of \clustervis is to interactively arrange the IP addresses (the circles). This is possible through 
 dividing the clusters based on different attributes of the log file or by creating own clusters. Communication behavior can be inspected interactively. 

\textbf{Visual Encoding} As the main visual paradigm we decided to use unit visualization \cite{drucker2015unifying,parkAtomGrammarUnit2018,huronVisualSedimentation2013}. 
These often combine the intuitiveness of unit visualizations with the appeal of physics-based animation and interactions, and are easy to learn for end users.
This paradigm supports all three requirements of the information security officer \textbf{R1-R3}.
The familiar character of the unit visualization provides insights into the firewall logs in an easily accessible way (\textbf{R1}). Representing the data as countable units also supports \textbf{T1.1}, namely to see the contained data. 
We decided for a cluster layout (\autoref{fig:use_case_1_clustervis} (a-e)) as the main view to facilitate an interactive organization of the presented data in a comprehensive form \textbf{T1.2}.
We chose the IP addresses to be represented by the units also with regard to \textbf{R2}, namely to get a picture of the current security situation within the scope of the recorded log. This includes \textbf{T2.1} to get an overview of the situation on the network and \textbf{T2.2} to see, which objects are active on the network and how they communicate with each other. 
Therefore, the amount of recorded connections for each IP address is encoded by the size of the circle. 
Connections between the IP addresses can be displayed on demand through links with arrows, resulting in a local node-link diagram.
We used the colors blue, yellow and white to encode whether an IP address is acting only as source, only as destination address, or as both.
Further anomalies can be displayed in red. This especially supports \textbf{T2.3}, allowing users to perceive the extent of the critical situation as well as locating the critical areas. 
The metaphorical unit visualization also supports \textbf{R3}, i.e. the communication of selected issues in a comprehensive way.
With the situation mode (\autoref{fig:use_case_1_clustervis} (f)) we added an additional layout to best support \textbf{R2}. Thereby the IP addresses are arranged in \textit{inside} and \textit{outside} with respect to the perimeter depending on their communication activity.
Additionally, to support inspection over time, a stacked bar chart is used showing the amount of active IP addresses over time. 

\begin{figure*}[t]
      \includegraphics[width=\linewidth]{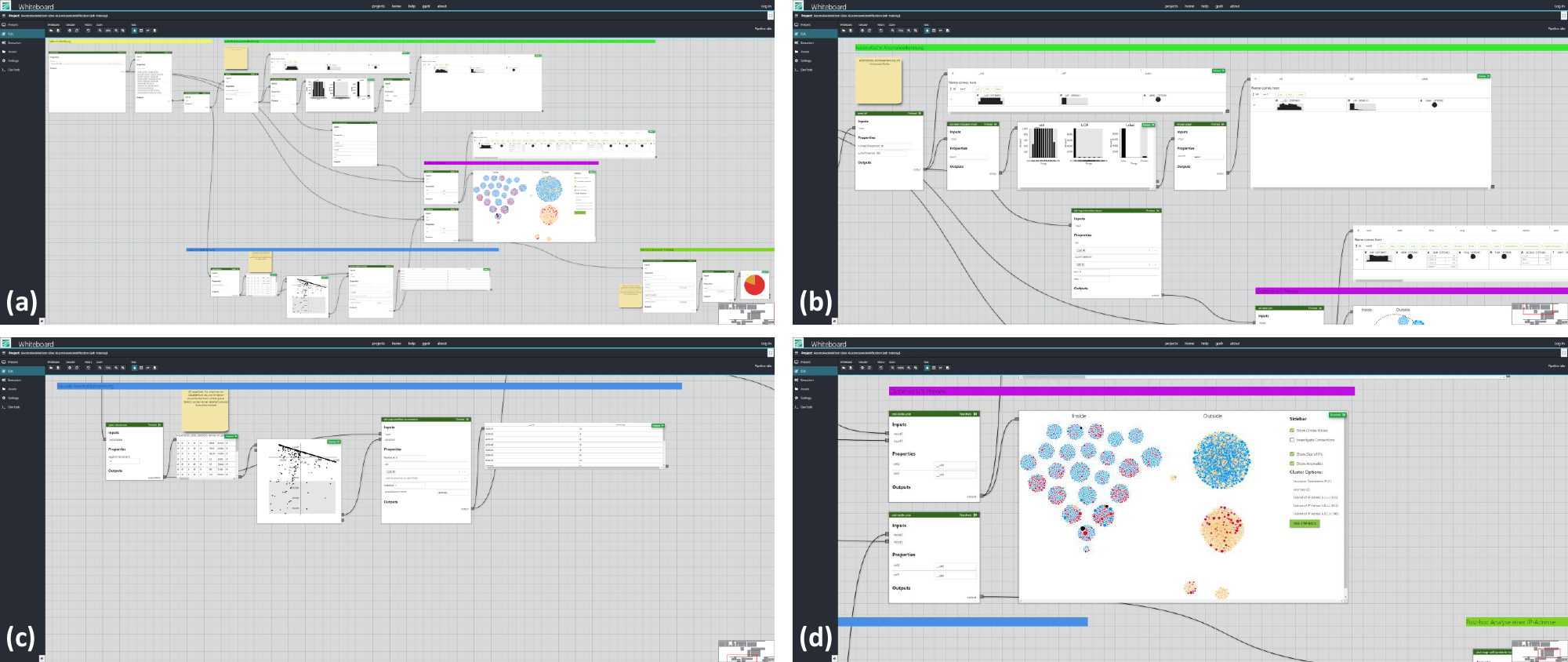}
      \caption{Usage scenario  - \textit{Whiteboard}: (a) Overview of the complete workflow built up with analysis blocks of \textit{Whiteboard}, consisting of 4 sections (color-coded): load data (yellow), inspect data (green), anomaly detection with LOF \cite{breunigLOFIdentifyingDensitybased2000}, \clustervis preview (violet), downstream analysis (dark green).
      (b) Automated part of the anomaly detection. A LOF algorithm is used and its results are displayed in various forms. (c) Manual and visual part of the anomaly detection including a scatterplot based on a PCA. (d) A simplified version of \clustervis is integrated in \whiteboard and can be used as a preview. Detected anomalies in the cluster are directly highlighted in red.
    }
      \label{fig:use_case_1_whiteboard}
  \end{figure*}

\textbf{Interactions}
To allow users to explore and organize the data with regard to personal needs \textbf{T1.2}, the main interaction provides means to cluster the circles according to different attributes. Selecting a cluster leads to a selection menu with attributes (\autoref{fig:use_case_1_clustervis} (b,c)).
Users can also create their own clusters and drag\&drop the circles from one cluster to another through direct manipulation \cite{hutchins1985direct}.
Further details and connections can be explored by selecting a particular circle (hovering/clicking) to support \textbf{T2.2}. Additional characteristics, such as "anomaly" can be added to the circle with regard to \textbf{T2.3}. The bar chart acts as a filter to select a time frame.
In the \textit{situation mode}, no clustering is supported, but users can explore the connections of the IP addresses through hovering and clicking on the circles.

\textbf{Data Abstraction \& Transformation}
For \clustervis we mainly pursued the following approaches to derive relevant information from the data:
\textit{Derive set of unique IP addresses:} Selecting all unique IP addresses from the source IP and destination IP attributes. \textit{Attributes for each IP address:} Summarizing the attributes of connections to "most common" for each IP address. \textit{Count amount of connections:} Counting the amount of entries for each IP address or for two communicating IP addresses. \textit{Amount of IP addresses over time:} Collecting the amount of IP addresses per time-frame based on the timestamp.

\textbf{Considering the psychological needs}
To satisfy the psychological need for \textit{stimulation}, the visualization is organized as an interactive playground. A force-based layout is used to arrange the units in circles. In this way, the layout is not predefined, supporting serendipity \cite{marchionini2006exploratory, leong2008choice, thudt2012bohemian} and an exploratory character (\textbf{R1}).
To address the psychological need for \textit{security}, we have designed an additional mode where the IP addresses are arranged in a more structured way, allowing the user to get a "picture of the situation" on the network (\textbf{R2}).
Following the approach in \cite{lenz2013exploring}, we selected interaction vocabulary based on the psychological needs and used them as inspiration for our design decisions.
For \textit{ClusterVis} we decided for the interaction vocabulary \textit{fast}, \textit{powerful} and \textit{spatial proximity} as we perceived that to be appropriate for the psychological need of \textit{stimulation}.
Additionally, to also support the psychological need of \textit{security} we decided for \textit{direct}, \textit{instant} and \textit{uniform}. 
Our decisions also goes along with the findings in \cite{lenz2013exploring}. 
Thus, the user can \textit{directly} interact with the units through direct manipulation \cite{hutchins1985direct}, namely by moving the clusters and units across the field. Direct manipulation also reflects \textit{spatial proximity}.
The movement \textit{instantly} follows the user's mouse movements and clicks. The cluster can be iteratively split into more clusters by  clicking on the cluster and selecting an attribute in the pop-up menue. This interaction stays the same (\textit{uniform}) for each level of clustering. 
Splitting a cluster results in a \textit{fast} and \textit{powerful} force-based division of the units.
To evoke the positive experience of \textit{popularity} as well as \textit{stimulation}, we designed for an appealing appearance.
This requirement was also explicitly stated by the stakeholders (especially \textbf{T1.1, T3.2, T3.3}) which led us to particularly focus on the visual aesthetics. 
We used circles for the units, as round objects are known to be especially aesthetically appealing \cite{Burch2015TheAO, ho2016influence,bar2006humans}. 
Further, we chose an aesthetically pleasing color scheme, fluid animations (force-based layout) and a lot of direct interaction (the user can perform many interactions by directly interacting with the visualized items). 
The main view is kept clean and simple, containing only the visualized and interactive data units.
We used a dark mode \cite{eisfeld2020rise, hakobyan2021impact}, which
makes the content of the visualization stand out more prominently.
However, as noticeable in \autoref{fig:use_case_1_whiteboard}, we do not enforce the dark mode and color selection. 

\subsection{Decisions for Whiteboard (Low-Level)}
The analytical part, called \textit{Whiteboard}, is a flow-based interface for interactive and flexible data analysis and is intended for the \textit{network analyst} (see \autoref{fig:requirements_tasks}). 
Analogous to a real-world whiteboard, users can place and connect analytical nodes on a white canvas and arrange them according to their needs. With the available data wrangling, machine learning and visualization nodes, the network analyst has a high degree of freedom to explore and inspect the firewall logs. 
While the development of \textit{Whiteboard} started before the provider's requirements for flexible data analysis became apparent, the requirements guided the further development of the tool.

\textbf{Visual Encoding}
To support \textbf{R4}, \textbf{R5} and \textbf{R6}, \whiteboard is designed to be an interface to create and execute data flows in a customized and flexible way. 
To support the flexibility (\textbf{T5.2}), the interface follows the interactive whiteboard metaphor.
The analysis steps are represented as rectangular interactive nodes and the data flows as lines connecting the nodes.
Domain experts can graphically create workflows by connecting executable nodes with each other to define the order of execution, visually resulting in a node-link diagram (see \autoref{fig:use_case_1_whiteboard}). 
 To support the inspection of the data analytically and visually (\textbf{T4.1}) and to zoom into suspicious spots (\textbf{T5.1}), \whiteboard contains multiple different predefined nodes for data processing  (e.g. filter - see e.g.\autoref{fig:use_case_1_whiteboard} (c)) and visualizations (e.g., table, bar chart, pie chart, or node-link diagram - see e.g. \autoref{fig:use_case_1_whiteboard} (b,c)), 
 and also for machine learning to support \textbf{T4.2}.
 Especially the visualization nodes are also well suited for the communication of issues to the supervisor (\textbf{T6.1}) as well as to new employees (\textbf{T6.2}) in an easily understandable way (\textbf{R6}).
 In particular, we integrated a version of \clustervis into \whiteboard (\autoref{fig:use_case_1_whiteboard} (a,d)).

\textbf{Interaction design}
To provide a high level of flexibility for the user (\textbf{T5.2}) the interface is based on the following basic interactions:
Users can select an area on the white space, create a new analysis node (block) and adjust the characteristics of the node directly at the node.
Second, users can draw connections between the nodes and adjust the layout by moving the nodes around, enabling \textbf{T4.2} and \textbf{T5.1}. They can also navigate over the whiteboard by zoom and pan. 
To fulfill \textbf{R4}, users can create a processing or machine learning pipeline, by creating and connecting appropriate blocks, such as visualizations or trainable anomaly detectors.
From intermediate results, subsets of data can be derived for more focused analysis (\textbf{R5}).
Finally, to address \textbf{R6} beyond other visualizations, user can add a node with the appealing \clustervis inside of \whiteboard to communicate the findings to the supervisor or colleagues.
Interactions with the visualizations, such as brushing, can be carried out in the nodes themselves and can be linked to other nodes by connecting them. Through this, \textbf{T5.1} and \textbf{T5.2} are well supported. 

\textbf{Data Abstraction \& Transformation}
As \whiteboard intrinsically provides functionalities to transform data, only transformation into a supported input format is required.
Further transformation depends on the goals of the analyst and is part of the interactive analysis process.
For the anomaly detection through machine learning, we used the original data table and used the log lines with selected attributes as input.
To use the internal \textit{ClusterVis}, the data table has to be transformed to match the expected input format.

\textbf{Considering the psychological needs}
The need for \textit{competence} is mainly addressed through the large collection of (more than 100) different node types, covering data processing, machine learning, and visualization, which the users can use as a toolkit to accomplish their goals.
The need for \textit{autonomy} is addressed through the flexible concept of the interface. 
For the \whiteboard the interaction vocabulary \textit{spatial proximity}, \textit{fluent}, and \textit{powerful} interactions over their opposites \textit{spatial separation}, \textit{stepwise}, and \textit{gentle} guided our decisions for the interaction design.
Following them, there are two ways how users can create new nodes: 
Either by clicking on the canvas and selecting the node of interest (\textit{spatial proximity}, \textit{powerful}) or by dragging the mouse from a node output to an empty space on the canvas (\textit{spatial proximity}, \textit{fluent}).  
\textit{Spatial proximity} also inspired the decision to visualize data in place, within the node (opposed to \textit{spatial separation}, as e.g. in RapidMiner \cite{mierswaYALERapidPrototyping2006}). 
Also the direct manipulation in the nodes (especially in the visualization nodes) reflects the principle of \textit{spatial proximity} well.

\section{Usage scenario}\label{sec:use cases}
With the usage scenario as presented in \autoref{fig:use_case_1_whiteboard} and \autoref{fig:use_case_1_clustervis} we want to demonstrate the potential for anomaly detection and exploration.
The task is to detect suspicious activities of anomalous IP addresses within a selected snippet of a firewall log.
\textbf{Whiteboard:}
The detection starts in \whiteboard (see \autoref{fig:use_case_1_whiteboard} a).
The analyst has constructed a pipeline based on the basic analytical elements on the canvas. The main workflow consists of loading the data set and processing the content through two analysis pipelines. The upper green pipeline (b) covers an automatic anomaly detection of abnormal log entries using the Local Outlier Factor (LOF) algorithm \cite{breunigLOFIdentifyingDensitybased2000} and extracts the identified outliers into a new list.
As the workflow is dynamic, it could be adjusted, extended or replaced anytime.
The lower blue pipeline (c) covers the interactive analysis of the data. The raw data is presented to the analyst in a 2D scatterplot after a PCA dimensionality reduction has been applied. Within the scatterplot node, the analyst can interactively decide which entries to select as outliers. The resulting selection is extracted into a new list, postprocessed to fit the data format requirements of \textit{ClusterVis}, which then displays the data and highlights anomalous IP addresses (see \autoref{fig:use_case_1_whiteboard} d).
As soon as the input data changes (e.g. the selection in the scatterplot), the downstream nodes will be updated with the new data automatically. 
\textbf{ClusterVis:}
The analyzed data log can then be exported as a log including a new column with the attribute \textit{Anomaly}. This export can be loaded into \clustervis for further exploration. \autoref{fig:use_case_1_clustervis} (a) shows the IP addresses that occur in the log during the selected time span. The anomalies are marked in red. 
Then, the anomalies can be separated from the rest by dividing the cluster accordingly (b,c). 
The anomalous values can, for example, be further divided according to the \textit{action} attribute (d). The anomalies in cluster \textit{accepted} might be of special interest and can be marked (green). Finally, the situation mode can be used to see the selected and anomalous IP addresses in a larger context (f). In this view, the IP addresses are arranged around the perimeter (firewall) as outside and inside the perimeter. IP addresses with a higher connectivity to the other side are positioned nearer to the perimeter.

  \begin{figure*}[t!]
     \includegraphics[width=\linewidth]{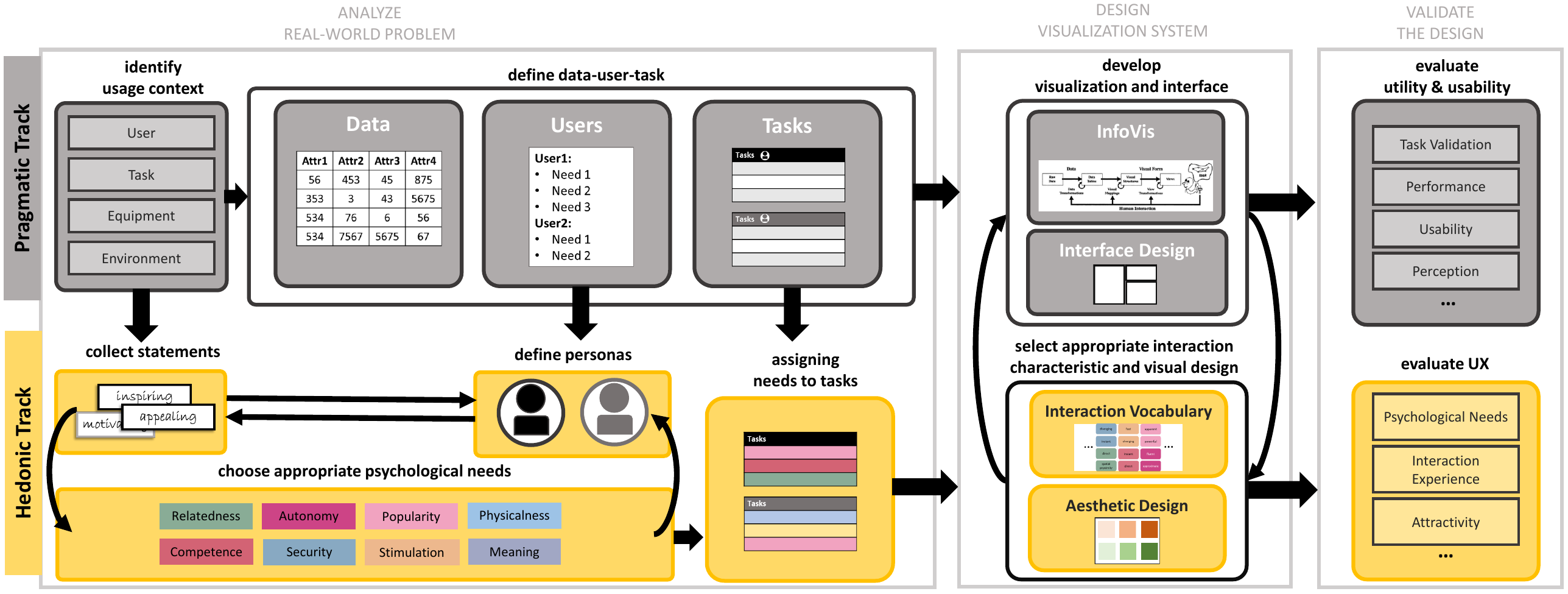}
      \caption{Design process model: Extension of design study process \cite{sedlmair2012design, card1999readings} by including a track taking into account hedonic qualities\cite{hassenzahl2000hedonic} through the incorporation of psychological needs\cite{hassenzahlNeedsAffectInteractive2010}, personas and interaction vocabulary \cite{lenz2013exploring}.
      }      
      \label{fig:proposed_design_process}
  \end{figure*}
\section{Wrap-up with information security officer} \label{sec:evaluation}
To validate our solution we conducted an observed experiment with the information security officer of our collaborating IT service provider. 
Reserving time with the security experts was difficult due to a high workload in this time frame.
However, we were glad to secure 4 hours with the information security officer, who has an excellent overview of the roles and tasks of various stakeholders in his organization. Therefore, he could give us valuable insights from the perspectives of management as well as network analysis. 
The core of the evaluation was a series of three prepared usage scenario with the real prototypes following a step-by-step description. For the usage scenario, real data of the organization was used. 
We observed the information security officer virtually through the shared screen. Thereby, he was free to express his thoughts and give comments. The wrap-up was conducted per video conference.
The three usage scenarios were: (1) detecting and highlighting an outlier in \clustervis and conducting a deeper analysis on the identified IP addresses in \textit{Whiteboard}; (2) identifying anomalies in \whiteboard with visual and with automated support and displaying them in \clustervis as critical areas; (3) detecting noticeable activities in \clustervis and analyzing them further within \textit{ClusterVis}.
We have designed the usage scenarios in such a way, that all of the tasks as defined in \autoref{fig:requirements_tasks} are covered.
Usage scenario (2) has been presented in\autoref{sec:use cases} and shown in \autoref{fig:use_case_1_whiteboard} and \autoref{fig:use_case_1_clustervis}.
The second important component of the evaluation was a structured interview, in which we asked whether the evaluated interfaces covered the requirements and tasks as listed in \autoref{fig:requirements_tasks} as well as specific questions regarding \textit{Witheboard}, \textit{ClusterVis} and the combination of the two interfaces in the overall system.

\textbf{Overall system:}
The information security officer \textit{agreed} that the combination of the two interfaces is enriching and helpful. He \textit{rather agreed} that it provides a good balance between an appealing possibility to communicate and a profound analysis. 
Here again, he rated possibility for \textbf{T 3.2} as \textit{neutral}. But he \textit{rather agreed} on \textbf{T 6.2}.
He also stated that he could imagine to use the system for the analysis of data other than firewall logs.
Finally, while he personally preferred the analytical \textit{Whiteboard}, he claimed that he would rather use the combination of both systems.
%
\textbf{Whiteboard:} 
After the assessment of the three usage scenarios, the information security officer \textit{rather agreed} that all but two tasks in \autoref{fig:requirements_tasks} can be supported by our prototypes.
One exception was the \textit{neutral} assessments for tasks \textbf{T 5.2} and \textbf{T 6.1}, where he was not able to assess that based on the guided usage scenarios.
He expressed the wish for a more high-level representation of the analysis results.
However, he \textit{rather agreed} that it was helpful for the identification of anomalies and patterns in the log data.
The implications regarding usability mainly addressed the reduction of complexity and the provision of ready-to-go pipelines and building blocks for specific common use cases. He also asked for an adaption of the interface's labels to the jargon of a network analyst. 
%
\textbf{ClusterVis:} Here again, the information security officer \textit{rather agreed} on the fulfillment of most requirements tasks listed in \autoref{fig:requirements_tasks} for the strategical perspective. One exception was \textbf{R2}
with a \textit{neutral} assessment. The information security officer also explicitly commented on all three sub-items of \textbf{R2}. He stressed that he can only \textit{rather agree} with the fulfillment of tasks that concern the analyst experts and those with a focus on the analysis of the specific firewall log.
Related to this, he has also rated \textbf{T3.2}
as \textit{rather disagree}.
For \textbf{T 3.3} the answer was between \textit{neutral} and \textit{rather agree} and dependent on the purpose and form in which the application would be reasonable. However, the application of \clustervis to communicate some information to the \textit{information center}  (and then to citizens) has been rated as rather likely. 
The security officer \textit{rather agreed}, that he found the interaction with \clustervis as particularly pleasant and aesthetically appealing and that the interaction  provided him with a good feeling.
Overall the interface was assessed as intuitive and usable. 
The main implication regarding the usability was to develop concepts to process and display larger data sets with a higher number of IP addresses covering larger time spans and, if possible, in real-time.  

\section{Reflection: Proposed Design Process}\label{sec:reflection}
In this section, we summarize how we integrated psychological needs and interaction characteristics \cite{hassenzahlNeedsAffectInteractive2010} into the design process for information visualization to strengthen the focus on user experience design. 
\autoref{fig:proposed_design_process} shows our proposed pipeline of an extended process model for design studies that takes these needs and characteristics into account.
It is divided into three parts, denoted by vertical blocks. This division is based on the first three steps of a design study as defined by Sedlmair et al. \cite{sedlmair2012design}, namely: \textit{analyze real-world problem}, \textit{design visualization system}, \textit{validate design}.
Our process contains a pragmatic track (depicted in gray), largely following the information visualization pipeline
on top and integrates UX design methods at the bottom (the  \textit{hedonic track}).
When comparing to the actual framework proposed by Sedlmair et al. \cite{sedlmair2012design}, our proposed design process refers to the \textit{core} design stage that consists of the steps \textit{discover}, \textit{design}, \textit{implement} and \textit{deploy}.

\paragraph{Analyze real-world problem}
This first block relates to the \textit{discover} stage \cite{sedlmair2012design}, where we extend the pragmatic steps of identifying the context of use \cite{ISO9241-UCD} and specifying the requirements according to data, user and task \cite{miksch2014matter} by additional hedonic methods. 
The first step of the pragmatic track is to identify the domain expert's problem or challenge. This includes the actual topic, the stakeholders, the requirements of the different stakeholders, and an overview of the available data.
In our case, this has been done mainly through conversations with the stakeholders and related research (see \autoref{sec:problem_characterization}). 
For the hedonic track, we want to motivate the inclusion of the following steps.
\textbf{Collect statements}
To better deduce the hedonic requirements, a collection of relevant phrases from the stakeholders can be gathered. These should be phrases expressing wishes for positive experience expressed by the stakeholders. 
Appropriate phrases can for example be captured during the interviews or by foraging the protocols afterwards. 
These phrases can further be used for decisions about psychological needs and for the definition of personas. 
\textbf{Choosing appropriate psychological needs}
To further support the empathy for the targeted UX, psychological needs can be taken into account. Considering psychological needs in the design process has two main advantages. On the one hand, they are helpful to better understand the users by taking into account their subjective preferences. On the other hand, they are helpful to make appropriate design decisions by "designing the experience" according to the so called be-goals \cite{hassenzahl2010experience}. Collecting relevant statements, choosing appropriate needs and defining personas all take place iteratively. Later, the selected needs are used to choose an appropriate interaction vocabulary. 
Different collections of the main psychological needs of a human exists, which can be used for this step (for example \cite{maslow1954motivation, sheldonWhatSatisfyingSatisfying2001a, reiss1998toward,flavell1986development}).
Further design frameworks from HCI, which incorperate these needs can be used as well \cite{hassenzahlDesigningMomentsMeaning2013, peters2018designing}.
We made use of the collection proposed by Hassenzahl et al. \cite{hassenzahlDesigningMomentsMeaning2013} containing 8 basic needs and the related tool \textit{need cards} \cite{hassenzahlExperienceDesignTools}. 
The illustrations and exemplary statements help to get a feeling for the needs.
\textbf{Define Personas}
To further support the empathy for the users during the design process, personas \cite{miaskiewicz2011personas} can be designed based on the information gathered about the targeted user group. Thereby, a group of multiple stakeholders can be summarized in one or multiple persona(s). For the definition and design of the personas both the collected statements and the psychological needs should be taken into account. In \autoref{fig:personas} we present an abstract version of the personas, including the collected statements and selected needs. However, during the design process we have developed more detailed persona descriptions following \cite{moser2013user}. 
\textbf{Assigning needs to tasks }
 A valuable step is to assign the selected psychological needs to the identified tasks either.
 Each task should be linked to the most appropriate need. 
This can help to define the tasks more precisely and identify tasks that are rather inappropriate for the targeted user. Further this can contribute to more accurate design decisions.
We recommend to apply the needs to the domain tasks, as the abstract visualization tasks contain less information about the user.
In hindsight and reflecting on our project, this step helped us refine the definition and grouping of the tasks and recalibrated our design goal.

\paragraph{Design visualization system}
This step relates to the stages \textit{design} and \textit{implement} in Sedlmair et al. \cite{sedlmair2012design}, including the generation and validation of data abstractions, visual encoding and interaction mechanisms. 
The pragmatic track includes the classical steps of the \textit{information visualization} process, as for example represented by the pipeline of Card et al. \cite{card1999readings} or the three inner layers of Munzner's nested model \cite{munzner2009nested}. 
We also considered \textit{user interface design} \cite{shneiderman2010designing} as an important building block in this step. 
Additionally, going along with other authors promoting a stronger integration of artistic design to increase the attractiveness or aesthetics in visualizations \cite{lau2007towards, moere2011role}, we argue to consider the following two aspects for the hedonic track.
\textbf{Interaction Vocabulary} 
In addition to the pragmatic rationale about appropriate interactions, a vocabulary with selected interaction characteristics \cite{lenz2013exploring} can be used with regard to the design for user experience \cite{hassenzahl2010experience}. 
Therefore, a set of interaction characteristics should be chosen for the targeted user experience. 
Here again, Hassenzahl et al. provides a tool, the \textit{interaction cards} \cite{hassenzahlExperienceDesignTools}.
According to Hassenzahl \cite{hassenzahl2010experience} there is no predefined way to select the interaction characteristics or to implement the characteristics. These choices are left to the designers and their creative ideas.  However, in Lenz et al. \cite{lenz2013exploring} some explanations for each attribute are provided, which can be used to map the interactions to the psychological needs. The characteristics can then be used as inspiration and reflection for each decision during the interaction design. The design process should be highly iterative and interlinked with the pragmatic interaction design.
In \autoref{sec:vis_design} we have shown how the selected characteristics influenced our design decisions.
\textbf{Aesthetic Design} 
As visual aesthetics are known to be important factors for users, on judgement about the appeal of a product \cite{hassenzahl2000hedonic}, \textit{aesthetic design} is the second crucial component of our hedonic track.
While design choices in InfoVis have been largely motivated by knowledge about human perception \cite{Ware12}, there are few guidelines for the use of visual aesthetics \cite{lau2007towards}.
At this step decisions on aesthetics (e.g. color harmony) have to be iteratively balanced with decisions from the pragmatic track (e.g. visual encoding). Related work that tries to incorporate aesthetics and artistic design to the domain of information visualization, as e.g. \cite{lau2007towards, moere2011role,samsel2018art}, can be used for deeper understanding.

\paragraph{Validate design}
This step can be related to the \textit{deploy} stage by Sedlmair et al. \cite{sedlmair2012design}, covering the validation and evaluation of the design. Beyond the question, how the validation has to be conducted in detail (for which there are other sources \cite{sedlmair2012design}, \cite{carpendale2008evaluating}, \cite{isenberg2008grounded}), 
to validate the design in accordance with our proposed design process, we suggest to enhance evaluation of the known pragmatic qualities of utility and usability with evaluation of the hedonic qualities.
\textbf{Evaluate user experience}
For the hedonic track it should be evaluated, whether the result meets the targeted psychological needs \cite{sheldonWhatSatisfyingSatisfying2001a}, how the character of the interactions is perceived \cite{lenz2013exploring} and also the overall perceived attractiveness (e.g., \cite{hassenzahl2003attrakdiff}) of the visualization.

\section{Discussion}\label{sec:discussion}
\textbf{Appropriateness of the solution:}
Our approach to design and implement separate but interlinked interfaces for each user group was perceived as adequate. 
The feedback revealed that our system would be particularly beneficial to support the operational activities around the analysis and planning of network-related issues.
For \whiteboard a promising direction would be to allow user-specific nodes in the form of ,,custom scripts'' or ,,external service integration".
\clustervis showed up as a conducive approach to facilitate the interactive exploration but lacks the capacity for large data sets.
\textbf{Generalizability of the approach:}
While we presented a solution based on the data and requirements of one particular organization, the solution is also applicable for other providers. 
Thereby, the particular role of the users might vary, as our personas represent a group of users with similar interests. For example,
a level-2 SOC analyst might also benefit from the interactive \textit{ClusterVis} to analyse the alerts prepared in \textit{Whiteboard}.
The information security officer also stated that he can well imagine to use the system even for other application areas and with other data. 
In fact, due to the high flexibility of \textit{Whiteboard}, the data input is not even constrained to network data.
\textit{ClusterVis} is also appropriate for other network data but can also be used for any data set with a unique identifier and an arbitrary number of data attributes. 
Moreover, a core functionality of \textit{Whiteboard} is the possibility to flexibly incorporate other visualizations than \textit{ClusterVis} to best suit a particular use case.
\textbf{Inclusion of hedonic user experience methods:}
We have found the inclusion of personas, psychological needs and interaction vocabulary as beneficial.
Being able to also include the affective notion of user statements helped us to keep more relevant context that informed our design process.
Therefore, the extended model explicitly includes a hedonic track as part of the design process.
However, evaluating the hedonic quality was challenging as the questions on emotions were quite unusual for the participant in the security context.
The helpful information on psychological needs and emotional states that we received from the interviews nevertheless encourages us to further explore the application of UX methods to information visualization design study.

\section{Conclusion}
In this paper, we have presented the process and the results of our design study on visualization solutions for firewall log analysis. Thereby, we have encountered two main clusters of interests and therefore have designed a solution consisting of two interlinked prototypes. While one targets the needs of the rather strategic group and exhibits a strong focus on appealing appearance, the other targets the operative group by focusing on flexible in-depth analysis. 
We showed how the prototypes evolved and were evaluated over time and also presented the feedback from a final assessment by our partner's information security officer. We also presented a usage scenario to validate the appropriateness of the solution.
We reflected on this design study with its diverse requirements for hedonic qualities 
by extending a visualization design process, widely used in the community, by a hedonic track. 
In future work we aim to refine the derived process model by applying it to our further projects and by examining possible extensions, e.g. incorporating other methods from user experience design.

 \acknowledgments{
This work was partly funded by the Hessian Ministry of the Interior and Sports (HMdIS) within the "Round Table Cybersecurity@Hessen" and by the German Federal Ministry of Education and Research and the Hessian Ministry of Higher Education, Research, Science and the Arts within their joint support of the National Research Center for Applied Cybersecurity ATHENE. We want to thank the Hessian Central Office for Data Processing (HZD) for the fruitful collaboration and feedback.
}

\bibliographystyle{abbrv-doi}
\bibliography{vizsec2022_firewalls}
\end{document}